\documentclass[epj]{svjour}
\newcommand{\bea}{\begin{eqnarray}}    
\newcommand{\eea}{\end{eqnarray}}      
\newcommand{\be}{\begin{equation}}
\newcommand{\ee}{\end{equation}}
\newcommand{\bef}{\begin{figue}}
\newcommand{\eef}{\end{figure}}

\newcommand{\cd}{{\langle n(r) \rangle_p}}

\def\spose#1{\hbox to 0pt{#1\hss}}
\def\ltapprox{\mathrel{\spose{\lower 3pt\hbox{$\mathchar"218$}}
\raise 2.0pt\hbox{$\mathchar"13C$}}}
\def\gtapprox{\mathrel{\spose{\lower 3pt\hbox{$\mathchar"218$}}
\raise 2.0pt\hbox{$\mathchar"13E$}}}
\def\inapprox{\mathrel{\spose{\lower 3pt\hbox{$\mathchar"218$}}
\raise 2.0pt\hbox{$\mathchar"232$}}}

\usepackage{graphics}
\begin{document}
\title{Statistical physics for cosmic structures}
\author{Francesco Sylos Labini\inst{1} \and Luciano Pietronero\inst{2}
}                     
\offprints{sylos@roma1.infn.it}          
\institute{
``Enrico Fermi Center'', Piazzale del Viminale 2,  00184 Rome, Italy \& 
``Istituto dei Sistemi Complessi'' CNR, 
Via dei Taurini 19, 00185 Rome, Italy \and 
Physics Department, University of Rome ``Sapienza'', 
Piazzale A. Moro 2, 00185 Rome, Italy 
}
\date{Received: date / Revised version: date}
%
\abstract{
The recent observations of galaxy and dark matter clumpy distributions
have provided new elements to the understanding of the problem of
cosmological structure formation.  The strong clumping
characterizing galaxy structures seems to be present in the overall
mass distribution and its relation to the highly isotropic Cosmic
Microwave Background Radiation represents a fundamental problem.  The
extension of structures, the formation of power-law correlations
characterizing the strongly clustered regime and the relation between
dark and visible matter are the key problems both from an
observational and a theoretical point of view. We discuss recent
progresses in the studies of structure formation by using concepts and
methods of statistical physics.
} 
\maketitle
\section{Introduction}
\label{intro}

In contemporary cosmological models the structures observed today at
large scales in the distribution of galaxies in the universe (see
Fig.1 --- discovered by the projects, e.g., 2dF \cite{2df}, SDSS
\cite{sdss,sdss2}) are explained by the dynamical evolution of purely
self-gravitating matter (dark matter) from an initial state with low
amplitude density fluctuations, the latter strongly constrained by
satellite observations of the fluctuations in the temperature of the
cosmic microwave background radiation (e.g. the satellites
COBE\cite{cobe} and WMAP\cite{wmap}). Despite the apparent simplicity
of the scheme, fundamental theoretical problems remain open and the
overall picture is based on the assumption that the main mass
component is dark.

\begin{figure} 
\resizebox{0.95\columnwidth}{!} {
\includegraphics{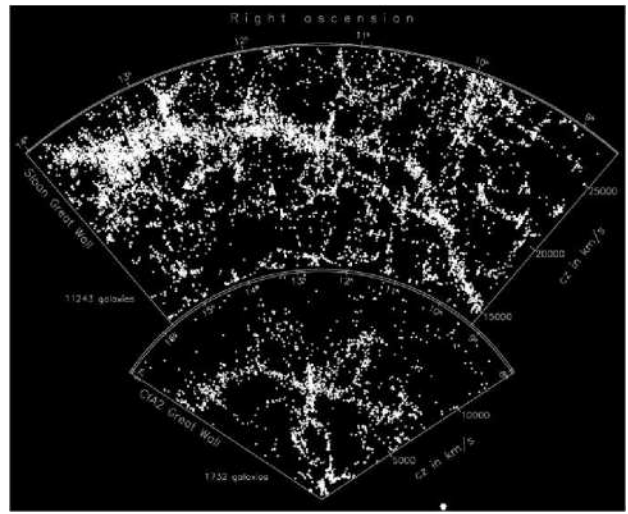}
}
\caption{Latest progress in redshift surveys.
SDSS Great Wall (2003) compared to CfA2 (1986) Great Wall at the same
scale. Redshift distances $cz$ are indicated. The small circle at the
bottom has a diameter of 5 Mpc/$h$, the clustering length according to
the standard interpretation of galaxy correlation.  The SDSS slice is
$4$ degrees wide, the CfA2 slice is $12$ degrees wide to make both
slices approximately the same physical width at the two walls. (From
\cite{book}).}
\label{figsdss}
\end{figure}

In this theoretical framework one crucial element is represented by
the initial conditions (IC) of the matter density field.  Models of
the early universe \cite{padm93} predict certain primordial
fluctuations in the matter density field, defining their correlation
properties and their relation to the present day matter
distribution. When gravity start to dominate the dynamical evolution
of density fluctuations, which can generally be described by the
Vlasov or ``collision-less Boltzmann'' equations coupled with the
Poisson equation, perturbations are still of very low amplitude.  One
of the most basic results (see e.g., \cite{pee80}) about
self-gravitating systems, treated using perturbative approaches to the
problem (i.e. the fluid limit), is that the amplitude of small
fluctuations grows monotonically in time, in a way which is
independent of the scale.  This linearized treatment breaks down at
any given scale when the relative fluctuation at the same scale
becomes of order unity, signaling the onset of the ``non-linear''
phase of gravitational collapse of the mass in regions of the
corresponding size.  If the initial velocity dispersion of particles
is small, non-linear structures start to develop at small scales first
and then the evolution becomes ``hierarchical'', i.e., structures build
up at successively larger scales. Given the finite time from the IC to
the present day, the development of non-linear structures is limited
in space, i.e., they can not be more extended than the scale at which
the linear approach predicts that the density contrast becomes of
order unity at the present time. This scale is fixed by the initial
amplitude of fluctuations, constrained by the CMBR, by the
hypothesized nature of the dominating dark matter component and its
correlation properties.

Observations of large scale galaxy distributions provide important
tests for these models. On the one hand the first question concerns
the extension of the regime of non-linear clustering and the intrinsic
properties of galaxy structures.  On the other hand according to this
scenario, at some large scales where fluctuations are still of small
amplitude, the imprints of primordial correlations should be preserved
and their detection represents a key observation for the validation of
the model.

In order to approach this complex problem, we use methods and concepts
of modern statistical physics \cite{book} to make a bridge between the
primordial fluctuation field and the development of large scale
structure in the universe. The first issue we discuss in what follows
concerns the correlation properties of the observed distribution of
galaxies and galaxy clusters, approaching this problem with the
perspective of a statistical physicist, exposed to the developments of
the last decades in the description of intrinsically irregular
structures, and by using instruments suitable to describe strong
irregularity, even if limited to a finite range of scales
\cite{slmp98,joyce05,nick2}. These methods offer a wider framework in
which to approach the problem of how to characterize the correlations
in galaxy distributions, without the a priori assumption of
homogeneity. That is, without the assumption that the distribution
inside a given sample is already uniform enough to give to a
sufficiently good approximation, the true (non-zero) mean density of
the underlying distribution of galaxies.  While this is a simple and
evident step for a statistical physicist, it can seem to be a radical
one for a cosmologist. After all the whole theoretical framework of
cosmology (i.e., the Friedmann-Robertson-Walker -- FRW -- solutions of
general relativity) is built on the assumption of an homogeneous and
isotropic distribution of matter. The approach we propose is thus an
empirical one, which surely is appropriate when faced with the
characterization of data. Further it is evidently important for the
formulation of theoretical explanations to understand and characterize
the data.

The second question in which the use of methods and concepts of
statistical physics allow us to clarify an important issue,
concerns the correlation properties of the initial matter density
fields in standard cosmological models.  In these models the matter
density field is described as having small fluctuations about a well
defined mean density and the initial conditions (i.e., very early in
the history of the universe) are specified by the so-called
Harrison-Zeldovich condition.  It is here that the concept of
``super-homogeneity'' introduced, for example, in the studies of
plasma and glass distributions, is relevant, as these models describe
fluctuations which are in fact of this type. Standard type models are
indeed characterized by {\it surface quadratic fluctuations} (of the
mass in spheres) and, for the particular form of primordial
cosmological spectra, by a negative power-law in the reduced
correlation function at large separations \cite{glass,lebo}. The
clarification of these properties, which correspond to a global
fine-tuning of positive and negative correlations, allow us to
define the strategy to measure such signals in real galaxy samples and
to identify several problems concerning, for example, the effects
related to sampling (galaxy distribution can be regarded as a sampling
of the underlying dark matter density field). 

The third issue in which a statistical physics approach maybe useful
concerns the theoretical modeling of non linear structure formation.
Analytical solutions of the Poisson-Vlasov equations are very
difficult to be formulated and the only instrument beyond the linear
regime is represented by numerical simulations. N-body simulations
solve numerically for the evolution of a system of $N$ particles
interacting purely through gravity, with a softening at very small
scales. The number of particles $N$ in the very largest current
simulations \cite{millenium} is $\sim 10^{10}$, many more than two
decades ago, but still many orders of magnitude fewer than the number
of real dark matter particles ($\sim 10^{80}$ in a comparable volume
for a typical candidate).  While such simulations constitute a very
powerful and essential tool, they lack the valuable guidance which a
fuller analytic understanding of the problem would provide.  The
question inevitably arises of the extent to which such numerical
simulations of a finite number of particles, reproduce the
mean-field/Vlasov limit of the cosmological models. The theoretical
questions concerns  the validity of this collisionless limit
and thus the crucial point is represented by the analysis of the
``discreteness effects'' \cite{prl,prd,sl,cg}.

As already mentioned, although dark matter is supposed to provide with
more than 0.9 of the total fraction of the mass-energy in universe
(see e.g. \cite{spergel03}), its amount and properties can only be
defined a posteriori. In addition the relation of dark matter to
visible matter is still not clear and the distribution itself of
visible matter requires more observations to be understood on the
relevant scales (see e.g. \cite{hogg04,joyce05}). More than twenty
years ago it has been surprisingly discovered that galaxy velocity
rotation curves remain flat at large distances from the galaxy center
while the density profile of luminous matters rapidly decays (see
e.g. \cite{rubin80}).  This is one of the strongest indications of the
need from dynamically dominant dark matter in the universe. Most
attention has been focused on the fact that these bound gravitational
systems contain large quantities of unseen matter and an intricate
paradigm has been developed in which {\it non-baryonic} dark matter
plays a central role not only in accounting for the dynamical mass of
galaxies and galaxy clusters but also for providing the initial seeds
which have given rise to the formation of structure via gravitational
collapse \cite{pee80}. In current standard cosmological models,
various forms of dark matter are needed to explain a number of
different phenomena, while baryons, which can be detected in the form
of, for example, luminous objects such as stars and galaxies, would
only be the $5\%$ of the total mass in the universe; the rest is made
of entities about which very little is understood: dark matter and
dark energy.  Very recently there have been developed observational
techniques which, by measuring the effect of gravitational lensing in
galaxy clusters \cite{lens}, or by measuring the gravitational
influence of structures on the CMBR \cite{rudnik}, are able to
reconstruct the three-dimensional distribution of dark matter and thus
allow a comprehension of the relative distribution of luminous and
dark matters, whose theoretical modeling is still lacking. These
observations have lead to surprising discoveries which rise new and
crucial questions to the validity of the standard interpretation of
structure formation \cite{newsci07}.


\section{Initial conditions and super-homogeneity}

The most prominent feature of the IC in the early universe, in
standard theoretical models, derived from inflationary mechanisms, is
that matter density field presents on large scale super-homogeneous
features~\cite{glass}.  This means the following. If one considers the
paradigm of uniform distributions, the Poisson process where particles
are placed completely randomly in space, the mass fluctuations in a
sphere of radius $R$ growths as $R^3$, i.e., like the volume of the
sphere. A super-homogeneous distribution is a system where the average
density is well defined (i.e., it is uniform) and where fluctuations in
a sphere grow slower than in the Poisson case, e.g., like $R^2$: in
this case there are the so-called surface fluctuations to
differentiate them from Poisson-like volume fluctuations.

A well known system in statistical physics systems of this kind is the
one component plasma~\cite{lebo} (OCP) which is characterized by a
dynamics which at thermal equilibrium gives rise to such
configurations. The OCP is simply a system of charged point particles
interacting through a repulsive $1/r$ potential, in a uniform
background which gives overall charge neutrality.  Simple
modifications of the OCP can produce equilibrium correlations of the
kind assumed in the cosmological context
\cite{lebo}.  

In terms of the  normalized mass variance 
$\sigma^2(R)=      
\langle M(R)^2 \rangle - \langle M(R) \rangle^2/ \langle M(R) \rangle^2$, 
where $\langle M(R) \rangle$ is the average mass in a sphere of radius
$R$ and $\langle M(R)^2 \rangle$ is the average of the square mass in
the same volume.  Thus for a Poisson distribution, where there are no
correlation between particles (or density fluctuations) at all, one
simply has $\sigma^2(R) \sim R^{-3}$. For an ordered system
characterized by small-scale anti-correlation the variance behaves as
$\sigma^2(R) \sim R^{-4}$ which is the fastest possible decay for
discrete or continuous distributions \cite{glass}.


The reason for this peculiar behavior of primordial density
fluctuations is the following.  In a FRW cosmology there is a
fundamental characteristic length scale, the horizon scale
$R_H(t)$. It is simply the distance light can travel from the Big Bang
singularity $t=0$ until any given time $t$ in the evolution of the
Universe, and it grows linearly with time.  The Harrison-Zeldovich
(H-Z) criterion can be written as $\sigma_M^2 (R=R_H(t)) = {\rm
constant}.$ This conditions states that the mass variance at the
horizon scale is constant: this can be expressed more conveniently in
terms of the power spectrum of density fluctuations
\cite{glass}
$P(\vec{k})=\left<|\delta_\rho(\vec{k})|^2\right>$
where $\delta_\rho(\vec{k})$ is the Fourier Transform of the
normalized fluctuation field $(\rho(\vec{r})-\rho_0)/\rho_0$, being
$\rho_0$ the average density. It is possible to show that
H-Z criterion  is equivalent to assume $P(k) \sim k$: in this
situation matter distribution present fluctuations of super-homogeneous type given
\cite{glass}.

The H-Z condition
is a consistency constraint in the framework of FRW cosmology. In fact
the FRW is a cosmological solution for a homogeneous Universe, about
which fluctuations represent an inhomogeneous perturbation: if density
fluctuations obey to a different condition than $P(k) \sim k$, then
the FRW description will always break down in the past or future, as
the amplitude of the perturbations become arbitrarily large or small.
For this reason the super-homogeneous nature of primordial density
field is a fundamental property independently on the nature of dark
matter. This is a very strong condition to impose, and it excludes
even Poisson processes ($P(k)=$ constant for small $k$)
\cite{glass}. 

Various models of primordial density fields differ for the behavior of
the power spectrum at large wave-lengths, i.e., at relatively small
scales \cite{padm93}. 
However at small $k$ they both exhibit the H-Z tail $P(k) \sim k$
which is in fact the common feature of all density fluctuations
compatible with FRW models.  Thus theoretical models of primordial
matter density fields in the expanding universe are characterized by a
single well-defined length scale, which is an imprint of the physics
of the early universe at the time of the decoupling between matter and
radiation \cite{padm93}.  The redshift characterizing the decoupling
is directly related to the scale at which the change of slope of the
power-spectrum of matter density fluctuations $P(k)$ occurs, i.e., it
defines the wave-number $k_c$ at which there is the turnover of the
power-spectrum between a regime, at large enough $k$, where it behaves
as a negative power-law of the wave number $P(k) \sim k^{m}$ with
$-1<m\le-3$, and a regime at small $k$ where $P(k)\sim k$ as predicted
by inflationary theories. Given the generality of this prediction, it
is clearly extremely important to look for this scale in the data. As
mentioned in the introduction the range of length-scales corresponding
to the regime of small fluctuations is linearly amplified during the
growth of gravitational instabilities. According to current models the
scales at which non-linear clustering occurs at the present time (of
order 10 Mpc) are much smaller than the scale $r_c$, corresponding to
the wave-number $k_c$, which is predicted to be $r_c \approx 124$
Mpc/h (where $0.5 <h<1$ is the normalized Hubble parameter) from
arguments based on CMBR anisotropies \cite{spergel03}. Thus the region
where the super-homogeneous features should still be in the linear
regime, allowing a direct test of the IC predicted by early universe
models.

At the scale $r_c$ the real space correlation function $\xi(r)$
(Fourier transform of the power spectrum) crosses zero, becoming
negative at larger scales. In particular the correlation function
presents a positive power-law behavior at scales $r \ll r_c$ and a
negative power-law behavior ($\xi(r) \sim - r^{-4}$) at scales $r \gg
r_c$. Positive and negative correlations are exactly balanced in way
such that the integral over the whole space of the correlation
function is equal to zero. This is a global condition on the system
fluctuations which corresponds to the fact that the distribution is
super-homogeneous.

By considering the observational features of super homogeneity one has
to take into account that in standard models galaxies result from a
{\it sampling} of the underlying dark matter density field: for
instance one selects (observationally) only the highest fluctuations
of the field which would represent the locations where galaxy will
eventually form. It has been shown that sampling a super-homogeneous
fluctuation field changes the nature of correlations~\cite{durrer},
introducing a stochastic noise which makes the system substantially
Poisson (e.g. $P(k) \sim$ constant) at large scales. However one may
show that the negative $\xi(r) \sim r^{-4}$ tail does not change under
sampling: on large enough scales, where in these models (anti)
correlations are small enough, the biased fluctuation field has a
correlation function which is linearly amplified with respect to the
underlying dark matter correlation function. For this reason the
detection of such a negative tail would be the main confirmation of
the super-homogeneous character of primordial density field
~\cite{book}.

The scale $r_c$ marks the maximum extension of positively correlated
structures: beyond $r_c$ the distribution must be anti-correlated
since the beginning, as there was no time to develop other
correlations.  The presence of structures, which mark long-range
correlations, whether or not of large amplitude, reported both by
observations of galaxy distributions (as those shown in Fig.1) and by
the indirect detection of dark matter 
\cite{lens,rudnik} is already pointing toward the fact that positive 
correlations extend well beyond $r_c$. For example, in \cite{rudnik}
it is shown that deep counts of radio-galaxies present a dip of about
$20-45 \%$ in the surface brightness at the location of a cold spot
observed in the CMBR anisotropies by the WMAP satellite. It is then
argued that if the cold spot does originate from structures at modest
redshift, to create, by gravitational interaction (the integrated
Sachs-Wolfe effect), the magnitude and angular size of the WMAP cold
spot it is required a $\sim$ 140 Mpc radius {\it completely} empty
void. This result, if confirmed, shows that (i) there are large-scale
structures of all matter (dark and visible) extended well beyond the
possible prediction of current models and that (ii) these structures
are of very large amplitude. This result must be tested in by the
analysis of three-dimensional galaxy catalogs. Up to now, measurements
of large samples of galaxy redshifts are not extended enough to reach
this region, where it is expected that $\xi(r) \sim -r^{-4}$, with the
appropriate and robust statistical properties. Future surveys, like
the complete SDSS catalog \cite{sdss}, may sample this range of
scales, but a precise study of the crossover to homogeneity,
discretization effects, sampling effects and statistical noise is
still required.

\section{Large scale galaxy distribution}

In the past twenty years observations have provided several three
dimensional maps of galaxy distribution, from which there is a growing
evidence of existence large scale structures.  This important
discovery has been possible thanks to the advent of large redshift
surveys: angular galaxy catalogs, considered in the past, are in fact
essentially smooth and structure-less. In the CfA2 catalog (1990)
\cite{cfa2}, which was one of the first maps surveying the local
universe, it has been surprisingly observed the giant ``Great Wall'',
a filament linking several groups and clusters of galaxies of
extension of about 200 Mpc/h. Recently the SDSS project \cite{sdss}
(2004---2009) has allowed to discover the ``Sloan Great Wall'' which
is almost double longer than the Great Wall. Nowadays this is the most
extended structure ever observed, covering about 400 Mpc/h, and whose
size is again limited by the boundaries of the sample. The search for
the ``maximum'' size of galaxy structures and voids, beyond which the
distribution becomes essentially smooth is still one of main open
problems. Instead it is well established that galaxy structures are
strongly irregular and form complex patterns.

The first question in this context concerns the studies of galaxy
correlation properties.  Two-point properties are particularly useful
to determine correlations and their spatial extension.  There are
different ways of measuring two-point properties and, in general, the
most suitable method depends on the type of correlation, strong or
weak, characterizing a given point distribution in a sample. 
The earliest observational studies, from angular catalogs,
produced the primary result \cite{tk69} that the {\it reduced
two-point correlation function} $\xi(r) \equiv \frac{\langle n(r)n(0)
\rangle}{\langle n \rangle^2}-1$ (where $n$ is the density of points)
is well approximated, in the range of scales from about $0.1$ Mpc/h to
$10$ Mpc/h, by a simple power-law: 
$\xi(r)\approx (r/r_0)^{-\gamma} $ with $\gamma \approx 1.8$ and
$r_0 \approx 4.7$ Mpc/$h$.  This result was subsequently confirmed by
numerous other authors in different redshift surveys (see e.g.,
\cite{davis88}). However, while $\xi(r)$ shows consistently a simple
power-law behavior characterized by this exponent, there is very
considerable variation among samples, with different depths and
luminosity cuts, in the measured {\it amplitude} of $\xi(r)$.  This
variation is usually ascribed {\it a posteriori} to an intrinsic
difference in the correlation properties of galaxies of different
luminosity (see e.g., \cite{davis88}):
brighter galaxies present larger values of $r_0$.  Theoretically it is
interpreted as a real physical phenomenon, as a manifestation of
``biasing''
\cite{kaiser84}.

Such a variation of the amplitude of the measured correlation function
may, however, be explained, entirely or partially, as a finite-size
effect i.e., as an artifact of statistical analysis in finite samples.
The explanation is as follows (see \cite{book}): The reduced
correlation function $\xi(r)$ can be written as $\xi(r)=\frac{\langle
n(r)\rangle_p}{\langle n \rangle}-1$, where ${\langle n(r)\rangle}_p$
is the {\it conditional density} of points, i.e., the mean density of
points in a spherical shell of radius $r$ centered on a galaxy.  The
latter is generally a very stable local quantity, the reliable
estimation of which at a given scale $r$ requires only a sample large
enough to allow a reasonable number of independent estimates of the
density in a shell. The mean density ${\langle n \rangle}$, on the
other hand, is a global quantity. The size of a sample in which it is
estimated reliably is not known {\it a priori}, but depends on the
properties of the underlying distribution.  Specifically the sample
must be large enough so that the mean density estimated in it has a
sufficiently small fluctuation with respect to the true asymptotic
average density.

It has been pointed out \cite{pie87} that, when analyzing a point
distribution which, like the galaxy distribution, is characterized by
large fluctuations, one should, in fact, first establish the existence
of a well defined mean density (and ultimately the scale at which it
becomes well defined and independent of the sample size, if it does)
before a statistic like $\xi(r)$, which measures fluctuations with
respect to such a mean density, is employed. Further the existence of
power-law correlations, which are clearly present in the galaxy
distribution, is typical of fractal distributions, which are
asymptotically empty.  In such distributions the mean density is
always strongly sample dependent, with an average value decreasing as
a function of sample size. Given the observation of such correlations
in the system, and the instability of the amplitude of the correlation
function $\xi(r)$ estimated in different samples, special care should
be taken in establishing first the scale (if any) at which homogeneity
becomes a good approximation.  The simplest way to do this is in fact
to measure the conditional density ${\langle n(r)\rangle}_p$. These
quantities are generally well defined, and give a characterization of
the two-point correlation properties of the distribution, irrespective
of whether the underlying distribution has a well defined mean density
or not.  A simple power law behavior $\cd = B r^{-\gamma}$ is
characteristic of scale-invariant fractal distributions, with the
exponent $\gamma<3$ giving the {\it fractal dimension} through
$D=3-\gamma$. The pre-factor $B$ is, in this case, simply related to
the lower cut-off of the distribution \cite{book}.  If the
distribution has a well defined mean density, one has, asymptotically,
${\langle n(r)\rangle}_p=$ constant $>0 $ (i.e., $D=3$ in the previous
formula). Measurements of this quantity can thus both characterize (i)
the regime of strong clustering and (ii) the scale and nature of a
transition to homogeneity.  Only once the existence of an average
density within the sample size is established in this manner does it
make sense to use $\xi(r)$.

Results in past catalogs (see \cite{book} and references therein) and
in preliminary samples of the SDSS
\cite{hogg04,nick2} show (Fig.2) 
\begin{figure} 
\resizebox{0.95\columnwidth}{!} {
\includegraphics{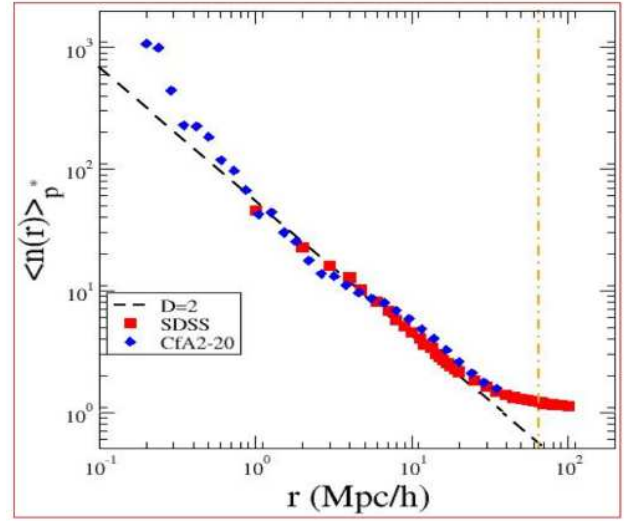}}
\caption{Behavior of the conditional density (red dots) 
in a preliminary sample of the SDSS survey \cite{hogg04}, together with
the determination of the conditional density (blue dots) in a sample
of the CfA2 catalog reported in \cite{joyce99}. There is a substantial
agreement between the two catalogs and that the new SDSS data seem to
show a flattening at about 70 Mpc/h.  A more detailed analysis is
required to study this transition and to characterize possible finite
size effects which may affect this behavior. (From \cite{joyce05})}
\label{figsdss}
\end{figure}
that in the range of scales [0.5,$\sim$ 30] Mpc/h galaxy distributions
are characterized by power-law correlations in the conditional density
in redshift space, with an exponent $\gamma=1.0
\pm 0.1$.  In the range of scales [$\sim$ 30,$\sim$ 100] Mpc/h there
are evidences for systematic unaveraged fluctuations corresponding to
the presence of large scale structures extending up to the boundaries
of the present survey, which require a detailed analysis of the
problems induced by finite volume effects on the determination of the
conditional density. In addition there are evidences which suggest
that in such range of scales the power-law index of the conditional
density has a smaller value. However future surveys will allow to
distinguish between the two possibilities: that a crossover to
homogeneity (corresponding to $\gamma=0$ in the conditional density)
occurs before 100 Mpc/h, or that correlations extend to scales of
order 100 Mpc/h (with a smaller exponent $0 < \gamma <1$).

Finally we note that even if a transition toward a constant value of
the conditional density will be finally detected this does not imply
that the distribution becomes uncorrelated on larger scales.  In fact,
this means that structures, beyond the crossover scale, have small
amplitude but they can be very well correlated on larger scales. It is
then in this situation where the detection of anti-correlations, which
as discussed above are predicted by all models of primordial density
fields, become the relevant issue to be addressed.

\section{Gravitational many-body problem} 

The understanding of the thermodynamics and dynamics of systems of
particles interacting only through their mutual Newtonian self gravity
is of fundamental importance in cosmology and astrophysics. In
statistical physics the problem of the evolution of self gravitating
classical bodies has been relatively neglected, primarily because of
the intrinsic difficulties associated with the attractive long-range
nature of gravity and its singular behavior at vanishing separation.
Long-range interacting systems (LRIS) present a series of peculiar
properties which make them qualitatively different from systems in
which the interactions between the component elements are
short-range. In the case of LRIS every element is coupled to every
other element in the system and not only with those located in a
finite neighborhood around itself. For this reason some of the most
basic concepts and instruments in physics, e.g. the
framework of equilibrium statistical mechanics, which have been
developed for short-range interacting systems, cannot be extended to
treat LRIS. One of the main feature of these systems is that
thermodynamical equilibrium is not generally reached.

Gravity is the paradigmatic example of LRIS and the peculiar features
of self gravitating systems have been mainly considered in the context
of astrophysics and cosmology. More recently \cite{ruffo} primarily
through the study of various simplified toy models, it has been shown
that LRIS generally exhibit a whole set of new qualitative properties
and behaviors: ensemble in-equivalence (negative specific heat,
temperature jumps), long-time relaxation (quasi-stationary states),
violations of ergodicity, subtleties in the relation of the fluid
(i.e., continuum) picture and the particle (granular) picture, etc..
These are commons to other physical laboratory systems such as systems
with unscreened Coulomb interactions and wave-particle systems
relevant to plasma physics \cite{ruffo}.

With the aim of approaching the problem of gravitational clustering in
the context of statistical mechanics it is natural to start by
reducing as much as possible the complexity of the analogous
cosmological problem and to focus on the essential aspects of the
problem. Thus we consider clustering without the expansion of the
universe, and starting from particularly simple initial conditions.
Our recent results suggest that in simplifying we do not loose any
essential elements which change the nature of gravitational
clustering \cite{prl,prd,sl,cg}.

The problem of the evolution of self gravitating classical bodies,
initially distributed very uniformly in infinite space, is as old as
Newton. Modern cosmology poses essentially the same problem as the
matter in the universe is now believed to consist predominantly of
almost purely self-gravitating particles which is, at early times,
indeed very close to uniformly distributed in the universe, and at
densities at which quantum effects are completely negligible. Despite
the age of the problem and the impressive advances of modern cosmology
in recent years, our understanding of it remains, however, very
incomplete.  In its essentials it is a simple well posed problem of
classical statistical mechanics.

\subsection{Discreteness effects in the linear regime}

We have recently formulated \cite{prl,prd} a perturbative theory of
the discrete N body problem which represents an useful approach to
control the problem of discreteness even in cosmological simulations
in the regime of small fluctuations, i.e., in the linear regime (see
Fig.3).  This situation is obtained by using as initial conditions of
the problem an infinite lattice of particles slightly displaced with
small or zero initial velocity dispersion. Thus up to a change in sign
in the force, the initial configuration is identical to the Coulomb
lattice (or Wigner crystal) in solid state physics (see
e.g. \cite{pines}), and we exploit this analogy to develop an
approximation to the evolution, in the linear regime, of the
gravitational problem.

More specifically, the equation of motion of particles moving under
their mutual self-gravity is \cite{nbs-standard-references}
\be
m_i {\ddot {\bf x} }_i 
= -
\sum_{i\neq j} 
\frac{G m_i m_j ({\bf x}_{i}-{\bf x}_j) }{|{\bf x}_{i}-{\bf x}_{j}|^3}\,.
\label{eom}
\ee
Here dots denote derivatives with respect to time $t$, ${\bf x}_i$ is
position of the $i$th particle of mass $m_i$.  We treat a system of
$N$ point particles, of equal mass $m$, initially placed on a Bravais
lattice, with periodic boundary conditions.  Perturbations from the
Coulomb lattice are described simply by Eq.~(\ref{eom}) with and $Gm^2
\rightarrow -e^2$ (where $e$ is the electronic charge).  As written in
Eq.~(\ref{eom}) the infinite sum giving the force on a particle is not
explicitly well defined. It is calculated by solving the Poisson
equation for the potential, with the mean mass density subtracted in
the source term. In the cosmological case this is appropriate as the
effect of the mean density is absorbed in the Hubble expansion; in the
case of the Coulomb lattice and of the gravitational static case
(which we consider here) it corresponds to the assumed presence of an
oppositely charged (negative mass for gravity) neutralizing background
(see discussion in
\cite{force}).

We consider now perturbations about the perfect lattice.
It is convenient to adopt the notation 
${\bf x}_i(t)={\bf R} + {\bf u}({\bf R},t)$
where ${\bf R}$ is the lattice vector of the $i$th particle, 
and ${\bf u}({\bf R},t)$ is the displacement of the particle from 
{\bf R}. Expanding to linear order in ${\bf u}({\bf R},t)$
about the equilibrium lattice configuration (in which the force on
each particle is exactly zero), we obtain
\be
{\bf {\ddot u}}({\bf R},t) 
= 
\sum_{{\bf R}'} 
{\cal D} ({\bf R}- {\bf R}') {\bf u}({\bf R}',t)\,. 
\label{linearised-eom}
\ee
The matrix ${\cal D}$ is known in solid
state physics, for any interaction, as the
{\it dynamical matrix} (see e.g. \cite{pines}). 
It is possible to compute the Fourier transform of  ${\cal D}$:
diagonalizing it one can determine, for each ${\bf k}$,  
three orthonormal eigenvectors ${\bf e}_n ({\bf k})$ and their
eigenvalues $\omega_n^2({\bf k})$ ($n=1,2,3$), which
obey \cite{pines} the Kohn sum rule 
$\sum_n \omega_n^2({\bf k}) = -4 \pi G \rho_0$,
where $\rho_0$ is the mean mass density.

At this point one may solve Eq.\ref{linearised-eom} by standard
techniques, obtaining that
$\tilde \delta({\bf k}) \sim \exp(\sqrt{4\pi G \epsilon_n({\bf k})t})$
where $\tilde \delta({\bf k})$ is the Fourier mode ${\bf k}$ of the
density contrast $\delta ({\bf r}) = (\rho ({\bf r}) - \rho_0)/ \rho_0$
and $\epsilon_n({\bf k}) = -\frac{\omega_n^2({\bf k})}{4 \pi G
\rho_0}$.  The eigenvalues are represented in Fig.3 (right panel) for
the case of a simple cubic lattice: we note that this particular case
presents both oscillating modes ($\epsilon_n({\bf k}) < 0 $) and modes
which grow faster ($\epsilon_n({\bf k}) >1$) than in the fluid limit
(which corresponds to $\epsilon_n({\bf k}) = 1 \; \forall {\bf k}$).

In the limit that     the  initial perturbations are  restricted    to
wavelengths much larger  than the    lattice spacing, the    evolution
corresponds exactly to that derived from an analogous linearization of
the  dynamics of  a pressure-less  self-gravitating  fluid.  Our  less
restricted approximation  allows one  to  trace the  evolution  of the
fully discrete distribution until the time when particles approach one
another, with modifications of the fluid limit explicitly depending on
the   lattice   spacing. Thus one   can   understand  exhaustively the
modifications introduced,  at  a given time and   length scale, by the
finiteness of $N$.

\begin{figure} 
\resizebox{0.99 \columnwidth}{!} {
\includegraphics{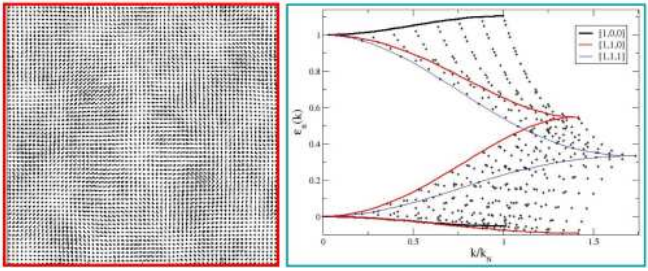}}
\caption{Initial 
condition for a N-body simulation corresponding to a perturbed lattice (left). 
In this situation density perturbations are small and a linear
analysis of the discrete problem allows one to identify a spectrum of
eigen-values (right) corresponding to different time scales of
collapse for the various wave-length of the perturbations. In the
fluid limit the time scale is the same for all modes and, in these
units, equal to one. (From \cite{prl}).}
\label{figlinear}
\end{figure}

\subsection{Toward the understanding of non-linear regime}

In an infinite space, in which the initial fluctuations are non-zero
and finite at all scales, the collapse of larger and larger scales
will continue ad infinitum. The system can therefore never reach a
time independent state, thus never reaching a thermodynamic
equilibrium.  One of the important results from numerical simulations
of such systems in the context of cosmology is that the system
nevertheless reaches a kind of scaling regime, in which the temporal
evolution is equivalent to a rescaling of the spatial variables
\cite{nbs-standard-references}.  This spatio-temporal scaling relation is referred
to as ``self-similarity''.

The evolution from above mentioned shuffled lattice (SL) initial
conditions converges, after a sufficient time, to a ``self-similar''
behavior, in which the two-point correlation function obeys a simple
spatio-temporal scaling relation. The time dependence of the scaling
is in good agreement with that inferred from the linearized fluid
approximation. Between the time at which the first non-linear
correlations emerge in a given SL and the convergence to this
``self-similar'' behavior, there is a transient period of significant
duration. During this time, the two-point correlation function already
approximates well, at the observed non-linear scales, a
spatio-temporal scaling relation, but in which the temporal evolution
is faster than the asymptotic evolution.  This behavior can be
understood as an effect of discreteness, which leads to an initial
``lag'' of the temporal evolution at small scales. The non-linear
correlations when they first develop are very well accounted for
solely in terms of two-body correlations. This is naturally explained
in terms of the central role of nearest neighbor (NN) interaction in the
build-up of these first non-linear correlations
\cite{thierry}. This two-body phase extends to the time of onset of
the spatio-temporal scaling, and thus the asymptotic form of the
correlation function is already established to a good approximation at
this time.

This situation has lead us to consider the comparison of the evolution
of such a system and that of ``daughter'' coarse-grained (CG) particle
distributions \cite{cg} (see Fig.4).  These are sparser (i.e., lower
density) particle distributions, defined by a simple coarse-graining
procedure, which share the same large-scale mass fluctuations.  In the
{\it numerical simulations} the CG particle distributions are observed
to evolve to give, after a sufficient time, two-point correlation
properties which agree well, over the range of scales simulated, with
those in the original distribution. Indeed both the original system
and its coarse-grainings converge toward a simple dynamical scaling
(``self-similar'') behavior {\it with the same amplitude}. The
characteristic time required for the CG system to begin to reproduce
the clustering in the original particle distribution at scales {\it
below} the CG scale increases as the latter scale does. These
observations are all very much in line with the qualitative picture of
the evolution of clustering widely accepted in cosmology: the CG
distributions share the same fluctuations at large scales and it is
these initial fluctuations alone, to a very good approximation, which
determine the correlations which develop at smaller scales at later
times.

As discussed above once particles begin to fall on one another there
is a phase in which very significant non-linear correlations develop
due to interactions between NN pairs of particles. The {\it form} of
the two-point correlation function which develops in this phase is
very similar to that observed, in the same range of amplitude, in the
asymptotic scaling regime at later times \cite{thierry}.  Thus it
appears that it is always possible to choose a CG of the original
system, which reproduces quite well the non-linear correlations in the
original system with this ``early time'', explicitly discrete,
dynamics of ``macro-particles'' of the CG distribution.  This provides
a simple physical picture/dynamical model for the generation of the
non-linear correlation function in the relevant range

This finding is very different to any existing explanations of the
dynamics giving rise to non-linear correlations in N body simulations
in cosmology. In this context theoretical modeling invariably assumes
that the non-linear correlations observed in simulations in this range
should be understood in the framework of a continuum Vlasov limit, in
which a mean-field approximation of the gravitational field is
appropriate. Indeed the fact that self-similarity is observed, with a
behavior independent of the particle density, is usually taken as an
indication that such a continuum description is appropriate. Our model
is manifestly not of this type, a key element is the discrete NN
dynamics, while also consistent with the amplitudes of the correlation
function being independent of particle density.

\begin{figure} 
\resizebox{0.95\columnwidth}{!} {
\includegraphics{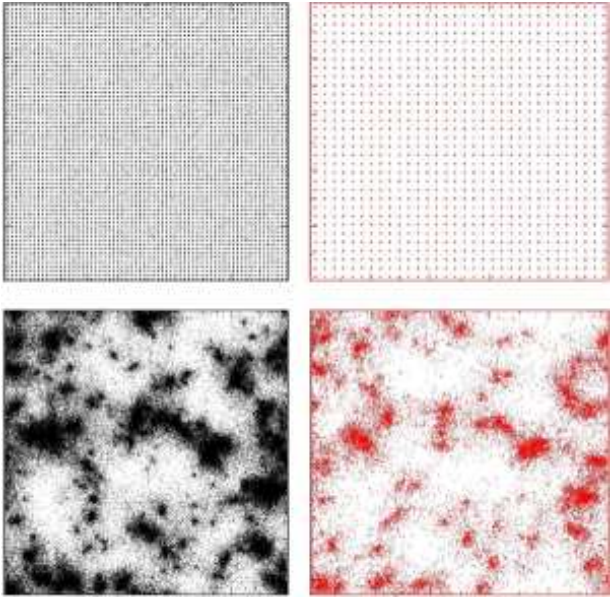}
}
\caption{Upper panels: Same 
initial conditions representing a randomly perturbed lattice,
with different number of points. Bottom panels: gravitationally
evolved systems. Despite the fact that the lower
resolution simulation has much less points, it traces the same
structures of the higher resolution one. The identification of the
similarities and differences among these systems allows one to
understand the effects related to the finiteness of the number of
points in the simulations. (From \cite{cg}).}
\label{figcg}
\end{figure}

\section{Conclusions}

The recent observations of galaxy and dark matter complex clumpy
distributions have provided new elements for the understanding of the
problem of cosmological structure formation.  The strong clumpiness
characterizing galaxy structures seems to be present in the overall
mass distribution and its relation to the highly isotropic CMBR
represents a fundamental problem.  In contemporary cosmological models
the structures observed today at large scales in the distribution of
galaxies are explained by the dynamical evolution of purely
self-gravitating matter from an initial state with low amplitude
density fluctuations.  The extension of structures, the formation of
power-law correlations characterizing the strongly clustered regime
and the relation between dark and visible matter are the key problems
both from an observational and a theoretical point of view.

In this puzzle statistical physics plays an important role in various
ways, which we have discussed above: (i) The complete characterization
of the correlations of visible and dark matter.  (ii) The analysis of
the very small anisotropies of the CMBR and their implications on the
initial fluctuations which recall the super-homogeneous properties
similar to plasmas and glasses.  (iii) The dynamical processes and
theories for the formation of complex structures from a very smooth
initial distribution and in a relatively short time.
\bigskip

It is a pleasure to thank Y. Baryshev, A. Gabrielli, M. Joyce,
B. Marcos, N. Vasilyev for useful collaborations and discussions.


\begin{thebibliography}{}

\bibitem{2df}  Colless M.M. et al., Mon.Not.R.Astron.Soc, {\bf 328},  (2001) 1039

\bibitem{sdss} York, D., et al., Astron.J., {\bf 120}, (2000) 1579 

\bibitem{sdss2} Adelman-McCarthy, J.K., et al  {\tt arXiv:0707.3413}

\bibitem{cobe}    Bennett, C.L., et al., 
Astrophys.J., {\bf 436},   (1994) 423

\bibitem{wmap}  Bennett, C.L., et al., 
Astrophys.J.Suppl., {\bf 148}, (2003)  1

\bibitem{padm93}
Padmanabhan, T., Phys. Rep.{\bf 188}, (1990) 285 



\bibitem{pee80} Peebles P.J.E., {\it Large Scale Structure of the Universe},        
(Princeton University  Press,  Princeton, New Jersey, 1980)  

\bibitem{book}  Gabrielli A., Sylos Labini F., Joyce M. and 
Pietronero L., {\it Statistical physics for cosmic structures } (Springer-Verlag, Berlin, 2005)

\bibitem{slmp98}  Sylos Labini, F., Montuori, M. \& Pietronero, L.,  
Phys.Rep., {\bf 293},  (1998) 66 

\bibitem{joyce05}
Joyce M., Sylos Labini F., Gabrielli A., Montuori M., Pietronero L., 
Astron.Astrophys. {\bf 443}, (2005)  11  

\bibitem{nick2}  Vasilyev N.L., Baryshev Yu. V., Sylos Labini F., Astron.Astrophys.
 {\bf 447}, 431 (2006)

\bibitem {glass}
Gabrielli, A., Joyce M. \& Sylos Labini F.,  Phys. Rev. {\bf D65},
   (2002) 083523

\bibitem{lebo}   Gabrielli  A.  Jancovici B, Joyce M.,  Lebowitz J., 
Pietronero L., Sylos Labini F.,   Phys.Rev., {\bf D67}, (2003)  043406

\bibitem{millenium} Springel V., et al.,   Nature, {\bf 435},(2005)  629 


\bibitem{prl}  Joyce M.,  Marcos B., Gabrielli A., Baertschiger T.,  Sylos Labini F., 
Physical Review Letters {\bf 95}, (2005) 011304 

\bibitem{prd}  Marcos B., Baertschiger T., Joyce M., Gabrielli A.,
Sylos Labini F.,  Phys.Rev. D {\bf 73}, (2006) 103507 

\bibitem{sl} Baertschiger T., Gabrielli A., Joyce, B. Marcos, 
F. Sylos Labini F., Phys.Rev.{\bf E 75},  (2007) 059905 

\bibitem{cg} Baertschiger T., Gabrielli A., Joyce, B. Marcos, 
F. Sylos Labini F.,   Phys.Rev.E in the press (2007) {\tt cond-mat/0612594}

\bibitem{spergel03} Spergel D.N. et al.,Astrophys.J. Suppl. ApJS, {\bf 148}, (2003) 175

\bibitem{hogg04}
Hogg D.W., et al., Astrophys.J., {\bf 624}, (2005) 54 



\bibitem{joyce99} Joyce, M., Montuori, M., Sylos Labini, F.,
Astrophys.J., {\bf 514}, (1999) L5 


\bibitem{rubin80} Rubin V. C., Thonnard N., Ford W. K., Astrophys.J, {\bf 238} (1980) 471


\bibitem{lens} R. Massey et al., Nature, {\bf 445}, (2007), 286

\bibitem{rudnik} L. Rudnick, S. Brown, L. R. Williams Astrophys.J. in the press 
(2007) {\tt arXiv:0704.0908v2}

\bibitem{newsci07}  A. Gefter, New Scientist {\bf 2594}, (2007) 30

\bibitem{durrer}
Durrer R., Gabrielli A., Joyce M., Sylos Labini F., Astrophys.J.{\bf
585}, (2003) L1

\bibitem{cfa2}  De Lapparent  V., Geller M. \& Huchra J.,   Astrophys.J.,  
{\bf 302}, (1986) L1

\bibitem{tk69} Totsuji H., Kihara T., Publ.Astron.Soc.Jpn,
{\bf 21}, (1969)  221

\bibitem{davis88} Davis, M. et al., Astrophys.J.Lett., {\bf 333}, 
 (1988) L9

\bibitem{kaiser84}  
Kaiser, N., 
Astrophys.J.Lett., {\bf 284},  (1984) L9


\bibitem{pie87} 
Pietronero, L.,  Physica A, {\bf 144},   (1987) 257

\bibitem{ruffo} Dauxois T., Ruffo S., Arimondo E., Wilkens M., eds.,
{\it Dynamics and Thermodynamics of Systems with Long-Range
  Interaction}, (Springer, Berlin 2002).

\bibitem{pines}
Pines D.,  {\it Elementary Excitations in Solids} 
(Benjamin, 1963). 

\bibitem{nbs-standard-references} 
Efstathiou G., et al., Astrophys. J. Supp. {\bf 57},  (1985) 241

\bibitem{force} Gabrielli A., Joyce M., Marcos B., Sylos Labini F.,
 Baertschiger T., Phys.Rev.{\bf  E74}  (2006) 021110 

\bibitem{thierry}
Baertschiger T. \& Sylos Labini F.,  
Phys.Rev.D,  {\bf 69}, (2004) 123001  

\end{thebibliography}
\end{document}